\documentclass[12pt]{article}
\usepackage[left=1in,right=1in,top=1in,bottom=1in]{geometry}
\usepackage{graphicx}
\usepackage{float}
\usepackage{cite}
\usepackage{amsfonts}
\usepackage{amssymb}
\usepackage{amsmath}
\usepackage{relsize}
\usepackage{xcolor}

\def\gtwid{\mathrel{\raise.3ex\hbox{$>$\kern-.75em\lower1ex\hbox{$\sim$}}}}
\def\ltwid{\mathrel{\raise.3ex\hbox{$<$\kern-.75em\lower1ex\hbox{$\sim$}}}}
\def\square{\kern1pt\vbox{\hrule height 1.2pt\hbox{\vrule width 1.2pt\hskip 3pt
   \vbox{\vskip 6pt}\hskip 3pt\vrule width 0.6pt}\hrule height 0.6pt}\kern1pt}

\begin{document}
\begin{titlepage}

\begin{flushright}
UFIFT-QG-23-05
\end{flushright}

\vskip 4cm

\begin{center}
{\bf Remembrance of Things Past}
\end{center}

\vskip 1cm

\begin{center}
 R. P. Woodard$^{\dagger}$ and B. Yesilyurt$^{*}$ 
\end{center}

\begin{center}
\it{Department of Physics, University of Florida, \\
Gainesville, FL 32611, UNITED STATES}
\end{center}

\vskip 1cm

\begin{center}
ABSTRACT
\end{center}

Nonlinear sigma models on de Sitter background have proved a useful prototype
for quantum gravity in summing the large logarithms which arise from loop
corrections. We consider a model whose evolution is described, at leading 
logarithm order, by the trace of the coincident, doubly differentiated scalar 
propagator. An analytic approximation for this quantity on an arbitrary 
expansion history is applied to generalize the resummed de Sitter result to any 
cosmological background which has experienced primordial inflation. In addition 
to analytic expressions, we present explicit numerical results for the evolution
in a plausible expansion history. The large scales of primordial inflation are 
transmitted to late times.

\begin{flushleft}
\end{flushleft}

\vspace{3cm}

\begin{flushleft}
$^{\dagger}$ e-mail: woodard@phys.ufl.edu \\
$^{*}$ e-mail: b.yesilyurt@ufl.edu 
\end{flushleft}

\end{titlepage}
\section{Introduction}

Many things change when quantum field theory is applied to a cosmological
background,
\begin{equation}
ds^2 = -dt^2 + a^2(t) d\vec{x} \!\cdot\! d\vec{x} \qquad \Longrightarrow \qquad
H(t) \equiv \frac{\dot{a}}{a} \quad , \quad \epsilon(t) \equiv -\frac{\dot{H}}{H^2}
\; . \label{geometry}
\end{equation}
One of these is the way the accelerated expansion ($\epsilon < 1$) of primordial 
inflation causes the energy-time uncertainty principle to predict the production 
of cosmological wavelength particles which are both massless and not conformally 
invariant \cite{Woodard:2014jba}. This is the basis for the primordial spectra 
of gravitons \cite{Starobinsky:1979ty} and scalars \cite{Mukhanov:1981xt}. 

Of course these particles must interact with themselves and with other quanta, 
which can change kinematics, long range forces, and even the expansion rate. 
Effects mediated by scalars are easier to compute, but model-dependent. Even on
the simplest background (de Sitter, with $\epsilon = 0$) graviton effects
require heroic computations  \cite{Tsamis:1996qk,Miao:2005am,Kahya:2007bc,
Leonard:2013xsa} and suffer from potential gauge dependence \cite{Garriga:2007zk,
Tsamis:2007is}, but they can do things which scalars cannot, and no one doubts 
that they can be described by general relativity as a low energy effective field 
\cite{Donoghue:1994dn,Donoghue:1995cz,Burgess:2003jk,Donoghue:2012zc,
Donoghue:2022eay}.

An important feature of loop corrections from inflationary gravitons is that 
they sometimes grow in time as more and more quanta are ripped from the vacuum. 
For example, a single loop of gravitons on de Sitter background changes the mode 
function of plane wave gravitons \cite{Tan:2021lza} and the Newtonian potential 
\cite{Tan:2022xpn} to,
\begin{eqnarray}
u(t,k) &\!\!\! = \!\!\!& u_0(t,k) \Biggl\{1 + \frac{16 G H^2}{3 \pi} \ln^2(a) 
+ O(G^2) \Biggr\} \; , \qquad \label{gravmode} \\
\Psi(t,r) &\!\!\! = \!\!\!& -\frac{G M}{a r} \Biggl\{1 + \frac{103 G}{15 \pi a^2 r^2}
- \frac{8 G H^2}{\pi} \ln^3(a) + O(G^2)\Biggr\} \; . \qquad \label{gravpot}
\end{eqnarray}
Similar results have been reported for 1-loop graviton corrections to fermions
\cite{Miao:2006gj,Miao:2012bj}, to massless, minimally coupled scalars 
\cite{Glavan:2021adm}, and to electromagnetism \cite{Glavan:2013jca,Wang:2014tza}.

It is fascinating to wonder what happens when the factors of $\ln[a(t)]$ grow
large enough to overwhelm the inflationary loop counting parameter of $G H^2$.
One would also like to generalize the background from de Sitter to an arbitrary
cosmological background (\ref{geometry}) in order to search for effects which
persist long after inflation has ended. Nonlinear sigma models provide a good
theoretical framework for such studies because they possess the same derivative
interactions as quantum gravity, without the complicated index structure and
the gauge problem \cite{Tsamis:2005hd,Kitamoto:2010et,Kitamoto:2011yx,
Kitamoto:2018dek}. A recent de Sitter background study \cite{Miao:2021gic,
Woodard:2023rqo} of this model,
\begin{equation}
\mathcal{L} = -\frac12 \partial_{\mu} A \partial_{\nu} A g^{\mu\nu} \sqrt{-g} -
\frac12 \Bigl(1 \!+\! \frac12 \lambda A\Bigr)^2 \partial_{\mu} B \partial_{\nu} B
g^{\mu\nu} \sqrt{-g} \; . \label{Lagrangian}
\end{equation}
showed that the large logarithms can be summed by combining a variant of 
Starobinksy's stochastic formalism \cite{Starobinsky:1986fx,Starobinsky:1994bd}
with a variant of the renormalization group. For example, the leading logarithms 
in the expectation value of the field $A(x)$ sum to the form,
\begin{equation}
\Bigl\langle \Omega \Bigl\vert A(x) \Bigr\vert \Omega \Bigr\rangle \longrightarrow
\frac{2}{\lambda} \Biggl[ \sqrt{1 + \frac{\lambda^2 H^2 \ln(a)}{16 \pi^2} } - 1
\Biggr] + \Bigl({\rm Stochastic\ acceleration}\Bigr) \; , \label{dSVEVA}
\end{equation}
where the final term speeds the field down its effective potential 
$V_{\rm eff}(A) = -\frac{3 H^4}{8 \pi^2} \ln\vert 1 + \frac12 \lambda A\vert$.

\newpage

The evolution of the scalar background (\ref{dSVEVA}) is entirely a stochastic
effect. One derives it by integrating out the differentiated $B$ fields in the 
presence of a constant $A$ background,
\begin{eqnarray}
\frac{\delta S[A,B]}{\delta A} &\!\!\! = \!\!\!& \partial_{\mu} \Bigl[ \sqrt{-g} 
\, g^{\mu\nu} \partial_{\nu} A\Bigr] - \frac{\lambda}{2} \Bigl(1 \!+\! 
\frac{\lambda}{2} A\Bigr) \partial_{\mu} B \partial_{\nu} B g^{\mu\nu} 
\sqrt{-g} \; , \qquad \\
&\!\!\! \longrightarrow \!\!\!& \partial_{\mu} \Bigl[ \sqrt{-g} \, g^{\mu\nu} 
\partial_{\nu} A\Bigr] - \frac{\lambda \sqrt{-g} \, g^{\mu\nu} \partial_{\mu} 
\partial'_{\nu} i\Delta(x;x') \vert_{x' = x}}{2 (1 + \frac{\lambda}{2} A)} = 0 \; . 
\qquad \label{stocheqn}
\end{eqnarray}
Because (\ref{stocheqn}) is the equation of motion for a scalar potential model,
one can sum its leading logarithms using Starobinsky's stochastic formalism
\cite{Starobinsky:1986fx,Starobinsky:1994bd}. The stochastic random field 
$\mathcal{A}(t,\vec{x})$ of this formalism obeys the Langevin equation,
\begin{equation}
-3 H(t) \Bigl[ \dot{\mathcal{A}}(t,\vec{x}) - \dot{\mathcal{A}}_0(t,\vec{x})
\Bigr] = \frac{\frac{\lambda}{2} \mathcal{T}[a](t)}{1 + \frac{\lambda}{2} 
\mathcal{A}(t,\vec{x})} \; , \label{Langevin}
\end{equation}
where the stochastic ``jitter'' $\mathcal{A}_0(t,\vec{x})$ is the infrared-truncated, 
free field mode sum,
\begin{equation}
\mathcal{A}_0(t,\vec{x}) \equiv \int_{a_i H_i}^{a H} \!\!\! \frac{d^3k}{(2\pi)^3} 
\sqrt{ \frac{H^2(t_k) C(\epsilon(t_k))}{2 k^3}} \Biggl\{ e^{-i \vec{k} \cdot \vec{x}} 
\alpha(\vec{k}) + e^{i \vec{k} \cdot \vec{x}} \alpha^{\dagger}(\vec{k}) \Biggr\} 
\quad , \quad a(t_k) H(t_k) \equiv k \; , \label{freefield}
\end{equation}
where inflation begins at time $t_i$ and $C(\epsilon) \equiv \frac1{\pi} 
\Gamma^2(\frac12 + \frac1{1-\epsilon}) [2(1 - \epsilon)]^{\frac{2}{1-\epsilon}}$. 
Here $\mathcal{T}[a](t)$ stands for the trace of the coincident, differentiated 
propagator which depends functionally on the expansion history $a(t)$,
\begin{equation}
\mathcal{T}[a](t) \equiv g^{\mu\nu} \partial_{\mu} \partial'_{\nu} i\Delta(x;x')
\Bigl\vert_{x'=x} \; . \label{taudef}
\end{equation}
Neglecting the stochastic jitter (which merely accelerates the rolling of 
$\mathcal{A}(t,\vec{x})$ down its potential) gives a first order equation which
can be integrated (from $\mathcal{A}(t_i,\vec{x}) = 0$) to give,
\begin{equation}
\mathcal{A}(t,\vec{x}) = \frac{2}{\lambda} \Biggl[ \sqrt{1 - \frac{\lambda^2}{6} 
\int_{t_i}^{t} \!\! dt' \frac{\mathcal{T}[a](t')}{H(t')} } \! - 1 \Biggr] + 
\Bigl({\rm Stochastic\ acceleration}\Bigr) \; . \label{genA}
\end{equation}

The de Sitter result (\ref{dSVEVA}) was obtained by using dimensional regularization
on de Sitter background to evaluate the trace of the coincident, differentiated 
propagator \cite{Onemli:2002hr,Onemli:2004mb},
\begin{equation}
\mathcal{T}\Bigl[{\rm de\ Sitter}\Bigr](t) = -\frac{H^D \Gamma(D)}{(4\pi)^{\frac{D}{2}} 
\Gamma(\frac{D}{2})} \longrightarrow - \frac{3 H^4}{8 \pi^2} \qquad \Bigl(D = 4\Bigr)
\; . \label{dStrace}
\end{equation} 
The purpose of this paper is to evaluate (\ref{genA}) using a recently developed 
analytic approximation of $\mathcal{T}[a](t)$ for a general expansion history
which has undergone primordial inflation \cite{Kasdagli:2023nzj}.\footnote{
Note also the related study of scalar perturbations amplified during
inflation to explain dark energy and the Hubble tension \cite{Glavan:2013mra, Glavan:2014uga, Glavan:2015cut, Glavan:2017jye, Belgacem:2021ieb, Vedder:2022spt}.} 
In section 2 we
review these results. In particular, we specify the renormalization condition, 
which was not needed for the de Sitter result (\ref{dStrace}). Section 3 devises a 
plausible expansion history and presents numerical results for (\ref{genA}) in this 
expansion history. Of course this requires us to make explicit choices for the 
coupling constant $\lambda$ and for the renormalization scale. Our conclusions
comprise section 4.

\section{Generalizing from de Sitter}

The purpose of this section is to review the analytic approximation developed in
\cite{Kasdagli:2023nzj} for the trace of the coincident, doubly differentiated 
scalar propagator (\ref{taudef}) on a general cosmological geometry (\ref{geometry})
which has undergone primordial inflation. We first give the primitive results,
both during inflation and afterwards. Then we implement renormalization.

\subsection{General Primitive Results}

Dolgov and Pelliccia have shown that $\mathcal{T}[a](t)$ can be inferred from the
coincidence limit of the scalar propagator $\mathcal{V}[a](t) \equiv i\Delta(x;x)$ 
\cite{Dolgov:2005se},
\begin{equation}
    \mathcal{T}[a](t) = -\frac{1}{2}\left( \frac{d}{dt} + (D-1)H \right)
    \Dot{ \mathcal{V}}[a](t) \ , \label{Tau}  
\end{equation}
where a dot indicates differentiation with respect to co-moving time $t$. One 
obtains $\mathcal{V}[a](t)$ for a general expansion history (which has undergone
primordial inflation) by first expressing it as a dimensionally
regulated, spatial Fourier mode sum of the norm-square of the mode function
$\mathcal{\Tilde{V}}(t,k) = u(t,k) u^*(t,k)$, 
\begin{equation}
    \mathcal{V}(t) =\int \!\! \frac{d^{D-1} k}{(2 \pi)^{D-1}} \, \mathcal{\Tilde{V}}(t,k) 
    = \frac{2}{\Gamma(\frac{D-1}{2}) (4 \pi )^{\frac{D-1}{2}}}\int_{a_i H_i}^{\infty} 
    \!\!\!\! dk \  k^{D-2} \mathcal{\Tilde{V}}(t,k)   \ . \label{modesV}
\end{equation}
We then employ appropriate analytic approximations for $\mathcal{\Tilde{V}}(t,k)$
depending upon the relation between the physical wave number $k/a(t)$ and the Hubble
parameter $H(t)$. During inflation $a(t) H(t)$ grows and there are two such regions:
\begin{itemize}
\item{{\it Ultraviolet} --- for $a(t) H(t) < k < \infty$; and}
\item{{\it Infrared} --- for $a(t_i) H(t_i) < k < a(t) H(t)$.}
\end{itemize}
\noindent Only the ultraviolet modes require dimensional regularization. When the
appropriate approximations are inserted in the mode sum the result during inflation
consists of a divergent part and a finite part \cite{Kasdagli:2023nzj},
\begin{eqnarray}
\mathcal{V}_{\mathrm{div}} &\!\!\! = \!\!\!& -\frac{1}{8}\left(\frac{D-2}{D-4}\right) 
\frac{[D-2 \epsilon(t)] H^{D-2}(t)}{\Gamma\left(\frac{D-1}{2}\right)(4 \pi)^{\frac{D-1}{2}}} 
\ , \qquad \label{divcont} \\
\mathcal{V}_{\mathrm{fin}} &\!\!\! = \!\!\!& -\frac{H^2(t)}{8 \pi^2} + \frac{1}{4 \pi^2} 
\int_{t_i}^t \!\! dt^{\prime} H^3\left(t^{\prime}\right) 
\left[1-\epsilon\left(t^{\prime}\right)\right] C\left(\epsilon\left(t^{\prime}\right)\right) 
\ . \qquad \label{infin}
\end{eqnarray}
Note that the horizon crossing relation $k = a(t_k) H(t_k)$ has been used to change
variables from $\frac{dk}{k}$ to $[1-\epsilon(t)] H(t) dt$. Recall that inflation
begins at time $t_i$, and the function $C(\epsilon)$ was defined after equation
(\ref{freefield}). Finally, note that the key derivative for (\ref{Tau}) is,
\begin{equation}
    \Dot{\mathcal{V}}_{\mathrm{fin}}= \frac{H^3}{4 \pi^2} 
    \Bigl[ 1-(1-\epsilon) [1 - C(\epsilon)] \Bigr] \ .  \label{infdotfin}
\end{equation}

Inflation ends at co-moving time $t_e$, after which $a(t) H(t)$ falls and some of
the infrared modes experience 2nd horizon crossing. This partitions the mode sum
into three regions:
\begin{itemize}
\item{{\it Ultraviolet} --- for $a(t_e) H(t_e) < k < \infty$;}
\item{{\it Near Infrared} --- for $a(t) H(t) < k < a(t_e) H(t_e)$; and}
\item{{\it Far Infrared} --- for $a(t_i) H(t_i) < k < a(t) H(t)$.}
\end{itemize}
If a mode experiences 1st horizon crossing at time $t$, then $t_2(t)$ represents
the time at which it experiences 2nd crossing. Similarly, if a mode experiences
2nd crossing at time $t$ then $t_1(t)$ represents the time at which it experiences
1st horizon crossing. Of course the ultraviolet divergence does not change but the
finite part after the end of inflation becomes \cite{Kasdagli:2023nzj},
\begin{equation}
    \begin{aligned}
& \mathcal{V}_{\mathrm{fin}}=\frac{(2-\epsilon) H^2}{8 \pi^2} 
\ln\left(\frac{k_e}{a H}\right) - \frac{H^2}{8 \pi^2} \left(\frac{k_e}{a H}\right)^2 \\
& +\frac{1}{4 \pi^2 a^2(t)} \int_{t_e}^t \!\!\! dt^{\prime} 
\left[\epsilon\left(t^{\prime}\right)-1\right] H\left(t^{\prime}\right) 
a^2\left(t^{\prime}\right) \times \cos^2\left[\int_{t_2\left(t_k\right)}^t \!\!\!\!
dt^{\prime \prime} \frac{k}{a\left(t^{\prime \prime}\right)}\right] \times H^2\left(t_1\right) C\left(\epsilon\left(t_1\right)\right) \\
& +\frac{1}{4 \pi^2} \int_t^{t_2\left(t_i\right)} \!\!\!\! dt^{\prime}
\left[\epsilon\left(t^{\prime}\right)-1\right] H\left(t^{\prime}\right) 
\times H^2\left(t_1\right) C\left(\epsilon\left(t_1\right)\right) \ . \label{ainfin}
\end{aligned}
\end{equation}
The factor $\cos^2\left[\int_{t_2\left(t_k\right)}^t dt^{\prime \prime} 
\frac{k}{a\left(t^{\prime \prime}\right)}\right]$ oscillates wildly and averages 
to 1/2 inside the integral. The key derivative for (\ref{Tau}) after inflation is,
\begin{eqnarray}
    \begin{aligned}
& \Dot{\mathcal{V}}_{\mathrm{fin}} = -\frac{\left[2 \epsilon(2-\epsilon)
+ \frac{1}{H} \Dot{\epsilon} \right] H^3}{8 \pi^2} \ln\left(\frac{k_e}{a H}\right)
- \frac{(1-\epsilon)(2-\epsilon) H^3}{8 \pi^2} + \frac{H^3}{4 \pi^2}
\left(\frac{k_e}{a H}\right)^2 \\
& -\frac{H(t)}{4 \pi^2 a^2(t)} \int_{t_e}^t \!\! dt^{\prime}
\left[\epsilon\left(t^{\prime}\right)-1\right] H\left(t^{\prime}\right) 
a^2\left(t^{\prime}\right) \times H^2\left(t_1\right) 
C\left(\epsilon\left(t_1\right)\right) \ . \ \label{afterdotfin}
\end{aligned}
\end{eqnarray}

\subsection{Renormalization}

The coincident field $A^2(x)$ is a composite operator and we must use composite 
operator renormalization to remove the divergence (\ref{divcont}) which afflicts
it \cite{Itzykson:1980rh,Weinberg:1996kr}. The only local, dimension two operator
with which $A^2$ can mix is the Ricci scalar $R = (D-1) [D - 2\epsilon] H^2$. 
Hence the appropriate counterterm is,
\begin{equation}
\Delta \mathcal{V}[a](t) = \Bigl(\frac{D \!-\! 2}{8}\Bigr) 
\frac{[D \!-\! 2 \epsilon(t)] H^2(t)}{\Gamma\left(\frac{D-1}{2}\right)(4 \pi)^{\frac{D-1}{2}}} 
\Bigl(\frac{\mu^{D-4}}{D-4}\Bigr) \ , \label{deltaV} 
\end{equation}
where $\mu$ is the scale of dimensional regularization. Adding (\ref{deltaV}) to 
(\ref{divcont}) and taking the unregulated limit gives a finite residual,
\begin{equation}
\lim_{D \rightarrow 4} \Bigl[\mathcal{V}_{\rm div}[a](t) - \Delta \mathcal{V}[a](t)
\Bigr] = \frac{R(t)}{48 \pi^2} \ln\Bigl[ \frac{\mu}{H(t)}\Bigr] \equiv 
\mathcal{V}_{\rm res}[a](t) \; . \label{Vresidual} 
\end{equation}
It is useful to make the running dimensionless with the parameterization $\mu = 
e^{\alpha} \times H(t_e)$,
\begin{equation}
\mathcal{V}_{\rm res}[a](t) = \frac{R(t)}{48 \pi^2} \ln\Bigl[ \frac{H(t_e)}{H(t)} 
\Bigr] + \frac{\alpha R(t)}{48 \pi^2} \equiv \frac{R(t)}{48 \pi^2} 
\ln\Bigl[ \frac{H(t_e)}{H(t)} \Bigr] + \mathcal{V}_{\rm run}[a](t) \; . \label{Vrun}
\end{equation}
Combining the first term with $\mathcal{V}_{\rm fin}$ defines the renormalized result,
\begin{equation}
\mathcal{V}_{\rm ren}(t) \equiv \mathcal{V}_{\rm fin}[a](t) + \frac{R(t)}{48 \pi^2} 
\ln\Bigl[ \frac{H(t_e)}{H(t)} \Bigr] \; . \label{Vren}
\end{equation}

\section{A Plausible Expansion History}

The purpose of this section is to develop an explicit expansion history
so that equation (\ref{genA}) can be numerically evolved. We begin by 
grafting a simple model of scalar-driven inflation onto the $\Lambda$CDM
model of post-inflationary cosmology. Then a convenient dimensionless
formulation is introduced for the various independent and dependent
variables. Finally, explicit results are presented for the scalar background.

\subsection{Attaching Inflation to a Hot Big Bang}

Our cosmology begins with inflation supported by the potential
$V(\varphi)$ of a scalar inflaton $\varphi$,
\begin{equation}
\ddot{\varphi} + 3 H \dot{\varphi} + V'(\varphi) = 0 \qquad \Longrightarrow
\qquad \left\{\begin{matrix}
H^2 = \frac{8\pi G}{3} [\frac12 \dot{\varphi}^2 + V(\varphi)] \\
\epsilon_{\varphi} = \frac{3 \dot{\varphi}^2}{ \dot{\varphi}^2 + 2 
V(\varphi)} \end{matrix}\right\} \; .
\end{equation}
In order to exploit simple slow roll expressions for the purposes of 
estimating parameters we chose the quadratic model,
\begin{equation}
V(\varphi) = \frac{c^2 \varphi^2}{16 \pi G} \qquad , \qquad c = 7.1 
\times 10^{-6} \; . \label{potential}
\end{equation}
With initial value $\varphi(t_i) = 15/\sqrt{8\pi G}$ one gets about 56.8 
e-foldings of inflation. This model agrees with the observed scalar amplitude 
and spectral index \cite{Planck:2018jri} but it badly violates the increasingly 
tight bounds on the tensor-to-scalar ratio \cite{BICEP:2021xfz}. However, for 
the illustrative purposes of this study that should not matter. Indeed, a more 
realistic plateau potential \cite{Starobinsky:1980te} would increase the accuracy 
of our analytic approximation (\ref{infin}) \cite{Kasdagli:2023nzj}.

\begin{figure}[ht]
    \centering
    \includegraphics[scale= 0.65]{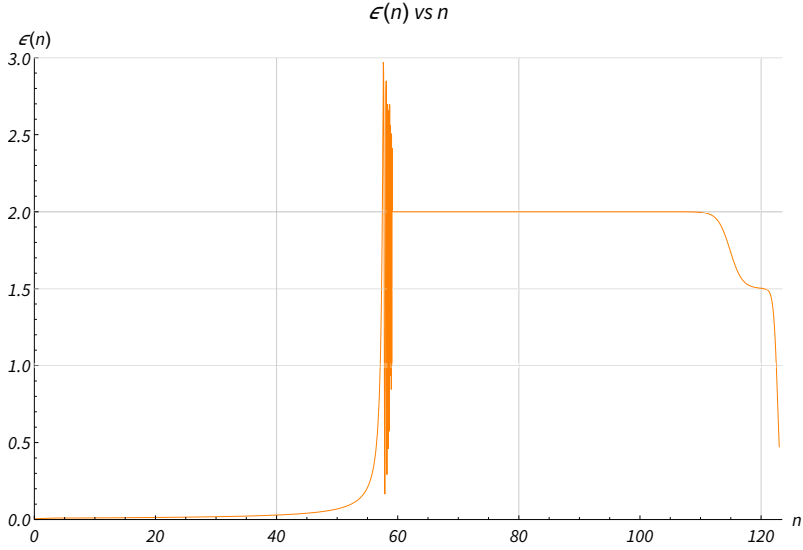}
    \caption{Evolution of the composite slow roll parameter (\ref{fulleps}) over
    cosmic history.}
    \label{fig:eps}
\end{figure}

Inflation ends when $\epsilon_{\varphi}(t_e) = 1$, after which $\epsilon_{\varphi}$ oscillates
between 0 and 3 with constant amplitude and increasing frequency. We want to 
graft this expansion history onto that of the $\Lambda$CDM model, 
\begin{equation}
\epsilon_{\Lambda}(t) = \frac{2 \Omega_r (1 \!+\! z)^4 + \frac{3}{2} 
\Omega_m (1 \!+\!z)^3}{\Omega_r (1 \!+\!z)^4 + \Omega_m (1 \!+\!z)^3 + 
\Omega_{\Lambda}} \qquad , \qquad 1 \!+\! z \equiv \frac{a(t_0)}{a(t)} \; . 
\label{LCDM} 
\end{equation}
where $t_0$ is the current time. The $\Lambda$CDM parameters are \cite{Planck:2018vyg},
\begin{equation}
\Omega_m \approx 0.315 \qquad , \qquad \Omega_{\Lambda} \approx 0.685 
\qquad , \qquad \Omega_r \approx \frac{\Omega_m}{3390} \; . \label{LCDMparams}
\end{equation}
Of course the $\Lambda$CDM model is not accurate at very early times when
the number of relativistic particles grows, but it will serve for our purpose
of providing a numerical framework to illustrate our analytic approximation
(\ref{ainfin}). Rather than devise an elaborate theory of reheating, we 
simply interpolate $\epsilon_{\varphi}(t)$ into $\epsilon_{\Lambda}(t)$,
\begin{equation}
\epsilon(t) \equiv \frac12 \Bigl[1 - \tanh(n \!-\! n_{\rm eq})\Bigr] \!\times\!
\epsilon_{\varphi}(t) + \frac12 \Bigl[1 + \tanh(n \!-\! n_{\rm eq})\Bigr]
\!\times\! \epsilon_{\Lambda}(t) \qquad , \qquad n \equiv \ln\Bigl[
\frac{a(t)}{a(t_i)}\Bigr] \; , \label{fulleps}
\end{equation}
where $n_{\rm eq} = 59.1$. Figure~\ref{fig:eps} shows the resulting first slow
roll parameter. 

\subsection{Dimensionless Formulation}

It is desirable to convert the evolution variable from the co-moving time $t$ 
to the dimensionless number of e-foldings since the beginning of inflation, $n \equiv
\ln[\frac{a(t)}{a(t_i)}]$. The resulting change in calculus is,
\begin{equation}
\frac{d}{d t} = H \frac{d}{d n} \qquad , \qquad \frac{d^2}{dt^2} = H^2 \Bigl[
\frac{d^2}{dn^2} - \epsilon \frac{d}{dn}\Bigr] \qquad , \qquad 
\int \!\! dt = \int \!\! \frac{dn}{H} \; . \label{calculus}
\end{equation}
Hence the relation (\ref{Tau}) between $\mathcal{T}[a](t)$ and $\mathcal{V}[a](t)$
becomes,
\begin{equation}
\mathcal{T}_{\rm ren} = -\frac{H^2}{2} \Bigl[\frac{d}{dn} + (3 \!-\! \epsilon) \Bigr]
\frac{d \mathcal{V}_{\rm ren}}{dn} \; . \label{Tefold}
\end{equation}
And the key integral in expression (\ref{genA}) can be written as,
\begin{equation}
\int_{t_i}^{t} \!\! dt' \frac{\mathcal{T}_{\rm ren}}{H(t')} = -\frac12 \!\! 
\int_{0}^{n} \!\! dn' \Bigl[ \frac{d}{dn'} + (3 \!-\! \epsilon)\Bigr] 
\frac{d \mathcal{V}_{\rm ren}}{dn'} \equiv - \frac{\mathcal{I}(n)}{16 \pi G} \; . 
\label{contI}
\end{equation}

\subsection{Numerical Results}

\begin{figure}[ht]
    \centering
    \includegraphics[scale= 0.75]{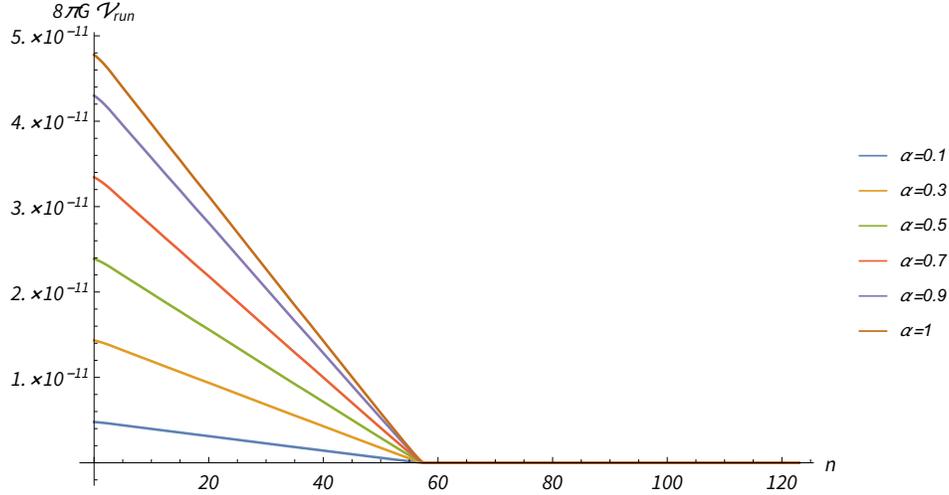}
    \caption{Evolution of ($8 \pi G$ times) the running contribution (\ref{Vrun}) 
    to the coincident propagator for different values of $\alpha$. Note that
    $\mathcal{V}_{\rm run}$ vanishes exactly during radiation domination, 
    and is too small to show up afterwards.}
    \label{fig:Alpha}
\end{figure}

\begin{figure}[H]
    \centering
    \includegraphics[scale= 0.6]{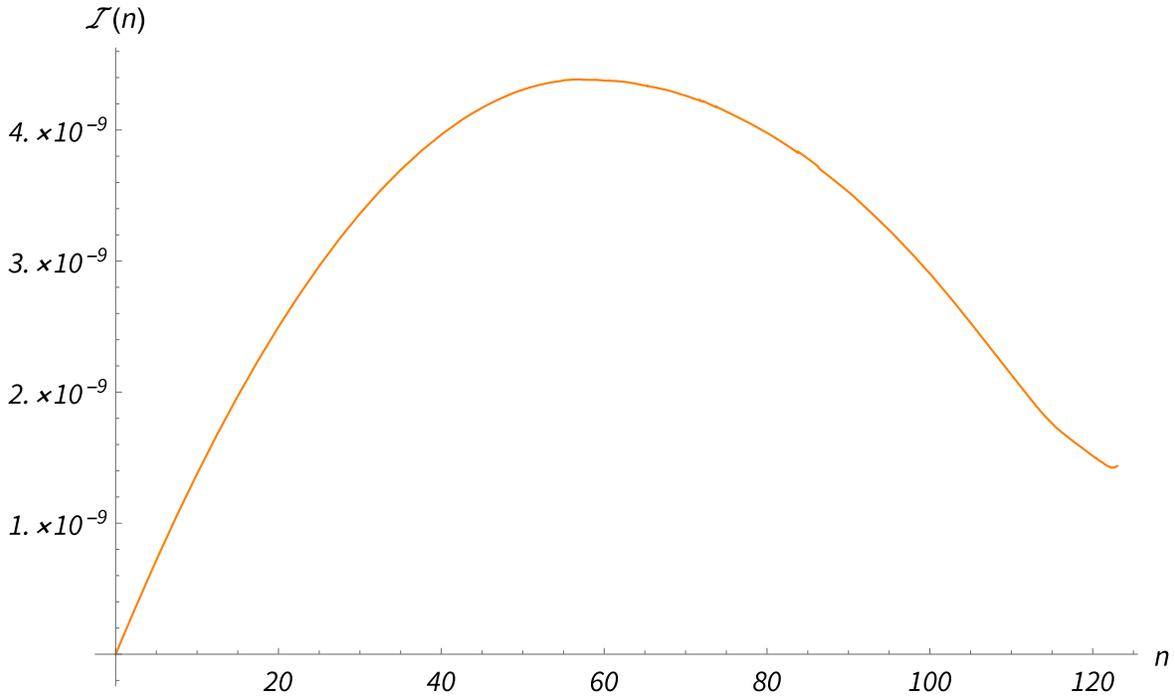}
    \caption{Evolution of $\mathcal{I}(n)$ from the beginning of primordial 
    inflation to today.}
    \label{fig:ivan}
\end{figure}

\begin{figure}[H]
    \centering
    \includegraphics[scale= 0.6]{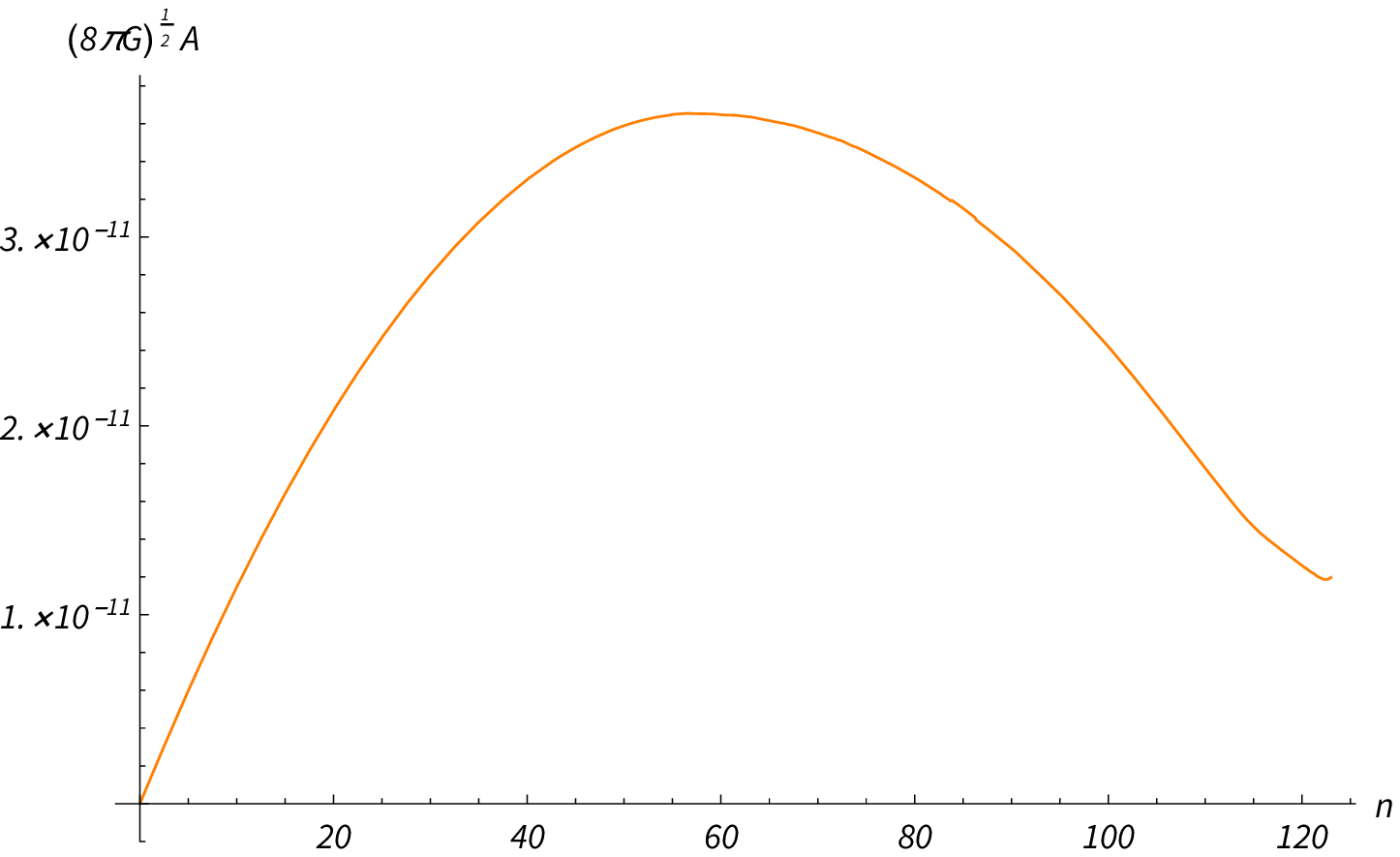}
    \caption{Evolution of the expectation value of $\sqrt{8 \pi G} \, A$ for 
    $\lambda = \frac1{10} \times  \sqrt{8 \pi G}$.}
    \label{fig:alyosha}
\end{figure}

As Figure~\ref{fig:Alpha} shows, the running term (\ref{Vrun}) makes only a small
contribution during inflation, and virtually nothing afterwards.
\noindent Figure~\ref{fig:ivan} shows the much larger contribution from the
dimensionless integral $\mathcal{I}(n)$ which was defined in expression (\ref{contI}).
Note that it grows during inflation, and then falls off afterwards. The expectation
value of the dimensionless field can be expressed in terms of $\mathcal{I}(n)$ as,
\begin{equation}
\Bigl\langle \Omega \Bigl\vert \sqrt{8 \pi G} \, A(x) \Bigr\vert \Omega \Bigr\rangle = 
\frac{2 \sqrt{8 \pi G}}{\lambda} \left[ \sqrt{1 \!+\! \frac{1}{12} 
\left( \frac{\lambda}{\sqrt{8\pi G}} \right)^2 \mathcal{I}(n)} \!-\! 1 \right] 
+ \Bigl({\rm Stochastic\ acceleration}\Bigr) \; . \label{vevAfin}
\end{equation}
\noindent Of course the right hand side of (\ref{vevAfin}) depends on our choice of 
the dimensionless coupling constant $\lambda/\sqrt{8 \pi G}$. Figure~\ref{fig:alyosha}
shows the analytic part for a convenient choice. We have ignored the ``stochastic 
acceleration'' term which increases the growth during inflation, and disappears 
afterwards. Even at the present day one can see that the dimensionless field retains 
more than half the value it built up during inflation. Had we employed a model with
a larger period of primordial inflation this effect would have been correspondingly
greater.

\begin{figure}[H]
    \centering
    \includegraphics[scale= 0.6]{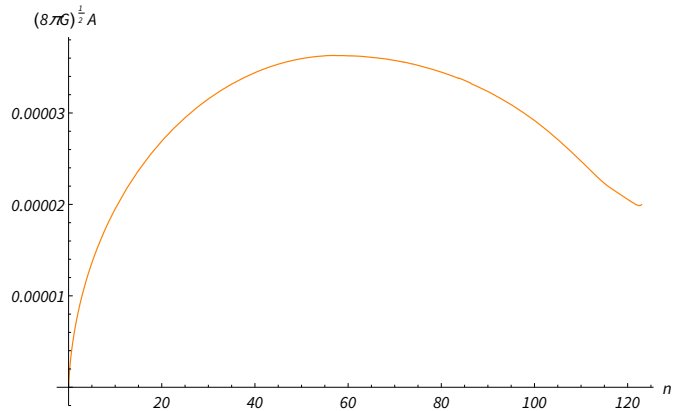}
    \caption{Evolution of the expectation value of $\sqrt{8 \pi G} \, A$ for 
    $\lambda = 10^6 \times  \sqrt{8 \pi G}$.}
    \label{fig:dmitri}
\end{figure}

One can see from Figure~\ref{fig:ivan} that $\mathcal{I}(n)$ is never larger than 
about $4 \times 10^{-9}$. This is because the inflationary model of section 2 persists 
for only the minimum of 50-60 e-foldings required to explain the Horizon Problem. One
can envisage much longer periods of inflation; extending the number of e-foldings by a 
factor increases the peak value of $\mathcal{I}(n)$ by roughly that same factor. 
Because the dimensionless coupling $\lambda/\sqrt{8 \pi G} = \frac1{10}$ chosen for 
Figure~\ref{fig:alyosha} is of order one, the square root in equation (\ref{vevAfin})
is near unity and perturbation theory never breaks down,
\begin{equation}
{\rm Perturbative} \qquad \Longrightarrow \qquad \Bigl\langle \Omega \Bigl\vert 
\sqrt{8 \pi G} \, A(x) \Bigr\vert \Omega \Bigr\rangle \simeq 
\frac{\lambda \mathcal{I}(n)}{12 \sqrt{8 \pi G}} \; . \label{pertA}
\end{equation}
However, expression (\ref{vevAfin}) is nonpertutbative, and we can make 
perturbation theory break down, either by increasing the duration of primordial
inflation or else by increasing the dimensionless coupling. The result in that case 
becomes,
\begin{equation}
{\rm Nonperturbative} \qquad \Longrightarrow \qquad \Bigl\langle \Omega \Bigl\vert 
\sqrt{8 \pi G} \, A(x) \Bigr\vert \Omega \Bigr\rangle \simeq 
\sqrt{\frac{\mathcal{I}(n)}{3}} \; . \label{nonpertA}
\end{equation}
Figure~\ref{fig:dmitri} illustrates this regime for $\lambda = 10^6 \times \sqrt{8 \pi G}$.

The alert reader will note from Figures~\ref{fig:ivan} through \ref{fig:dmitri}
that the decrease due to modes experiencing 2nd horizon crossing ceases at the
onset of the current phase of cosmic acceleration. Indeed, one might expect to
see the decline reversed because modes again begin exiting the horizon. However,
it must be recalled that the strength of the amplitude scales like the square of
the Hubble parameter at 1st crossing, as per equation (62) in Ref. 
\cite{Kasdagli:2023nzj}. Because the current Hubble parameter is some 55 orders
of magnitude smaller than its value during inflation, the predicted late time
growth would not be visible on our plots, even if they were extended beyond the
present time. 

\section{Conclusions}

We have analyzed the evolution of the expectation value of the scalar field $A$ 
in a non-linear sigma model (\ref{Lagrangian}) on a general cosmological background 
(\ref{geometry}) that experiences primordial inflation. We did this by solving the
Starobinsky Langevin equation (\ref{stocheqn}) for this theory \cite{Miao:2021gic}
which involves $\mathcal{T}[a](t)$, the trace of the doubly differentiated, 
coincident propagator (\ref{taudef}). Expression (\ref{Tau}) gives $\mathcal{T}[a](t)$
in terms of the coincident propagator $\mathcal{V}[a](t)$. A good analytic 
approximation \cite{Kasdagli:2023nzj} for the dimensionally regulated, primitive 
form of $\mathcal{V}[a](t)$ was reviewed in section 2.1, and subjected to composite
operator renormalization in section 2.2. The final result takes the form of a 
small ``running'' term proportional to the local Ricci scalar, plus a much larger
contribution (\ref{Vren}), with (\ref{infin}) and (\ref{infdotfin}) pertaining 
during inflation and (\ref{ainfin}) and (\ref{afterdotfin}) afterwards.

Although our analytic approximations should apply for any cosmology which
which has experienced primordial inflation, we devised an plausible expansion
history in section 3.1 to provide explicit numerical results. Figure~\ref{fig:eps}
shows the first slow roll parameter of this cosmology, from which the Hubble 
parameter and the scale factor can be constructed by integration. The small
running contribution (\ref{Vrun}) to $\mathcal{V}[a](t)$ is shown in 
Figure~\ref{fig:Alpha}. Figure~\ref{fig:ivan} gives the key integral (\ref{contI})
which determines the expectation value of the field shown in Figures~\ref{fig:alyosha}
and \ref{fig:dmitri}.

A major goal of this study was to quantify the extent to which the high scales 
of primordial inflation are transmitted to late times. In this regard it is worth
noting that the expectation value of $A$ would {\it vanish} if the Universe was
eternally radiation dominated (that is, $\epsilon(t) = 2$), or if it was flat.
The fact that we get nonzero results, even during radiation domination, is due
to the initial phase of primordial inflation. Note also from Figures~\ref{fig:alyosha}
and \ref{fig:dmitri} that, even at very late times, the field retains a substantial 
amount of the amplitude it built up during primordial inflation. Note finally that 
an arbitrarily large amplitude could be built up by extending the duration of 
primordial inflation, whereas there can be no extension of the post-inflationary 
diminution.

This study demonstrates that it is not only possible to sum the series of leading 
logarithms of nonlinear sigma models on de Sitter, but also on an arbitrary 
cosmological background (\ref{geometry}) which has undergone primordial inflation.
And we have seen two crucial things:
\begin{itemize}
\item{That evolution never stops; and}
\item{That an arbitrarily large effect from primordial inflation can be transmitted 
to late times.}
\end{itemize}
The next step is extending this formalism to quantum gravity in order to discover 
what becomes of graviton loop corrections such as (\ref{gravmode}-\ref{gravpot}) 
at late times and after perturbation theory has broken down.

\vskip 1cm

\centerline{\bf Acknowledgements}

This work was partially supported by NSF grant PHY-2207514 and by the
Institute for Fundamental Theory at the University of Florida.



\begin{thebibliography}{99}

\bibitem{Woodard:2014jba}
R.~P.~Woodard,
Int. J. Mod. Phys. D \textbf{23}, no.09, 1430020 (2014)
doi:10.1142/S0218271814300201
[arXiv:1407.4748 [gr-qc]].

\bibitem{Starobinsky:1979ty}
A.~A.~Starobinsky,
JETP Lett. \textbf{30}, 682-685 (1979)

\bibitem{Mukhanov:1981xt}
V.~F.~Mukhanov and G.~V.~Chibisov,
JETP Lett. \textbf{33}, 532-535 (1981)

\bibitem{Tsamis:1996qk}
N.~C.~Tsamis and R.~P.~Woodard,
Phys. Rev. D \textbf{54}, 2621-2639 (1996)
doi:10.1103/PhysRevD.54.2621
[arXiv:hep-ph/9602317 [hep-ph]].

\bibitem{Miao:2005am}
S.~P.~Miao and R.~P.~Woodard,
Class. Quant. Grav. \textbf{23}, 1721-1762 (2006)
doi:10.1088/0264-9381/23/5/016
[arXiv:gr-qc/0511140 [gr-qc]].

\bibitem{Kahya:2007bc}
E.~O.~Kahya and R.~P.~Woodard,
Phys. Rev. D \textbf{76}, 124005 (2007)
doi:10.1103/PhysRevD.76.124005
[arXiv:0709.0536 [gr-qc]].

\bibitem{Leonard:2013xsa}
K.~E.~Leonard and R.~P.~Woodard,
Class. Quant. Grav. \textbf{31}, 015010 (2014)
doi:10.1088/0264-9381/31/1/015010
[arXiv:1304.7265 [gr-qc]].

\bibitem{Garriga:2007zk}
J.~Garriga and T.~Tanaka,
Phys. Rev. D \textbf{77}, 024021 (2008)
doi:10.1103/PhysRevD.77.024021
[arXiv:0706.0295 [hep-th]].

\bibitem{Tsamis:2007is}
N.~C.~Tsamis and R.~P.~Woodard,
Phys. Rev. D \textbf{78}, 028501 (2008)
doi:10.1103/PhysRevD.78.028501
[arXiv:0708.2004 [hep-th]].

\bibitem{Donoghue:1994dn}
J.~F.~Donoghue,
Phys. Rev. D \textbf{50}, 3874-3888 (1994)
doi:10.1103/PhysRevD.50.3874
[arXiv:gr-qc/9405057 [gr-qc]].

\bibitem{Donoghue:1995cz}
J.~F.~Donoghue,
[arXiv:gr-qc/9512024 [gr-qc]].

\bibitem{Burgess:2003jk}
C.~P.~Burgess,
Living Rev. Rel. \textbf{7}, 5-56 (2004)
doi:10.12942/lrr-2004-5
[arXiv:gr-qc/0311082 [gr-qc]].

\bibitem{Donoghue:2012zc}
J.~F.~Donoghue,
AIP Conf. Proc. \textbf{1483}, no.1, 73-94 (2012)
doi:10.1063/1.4756964
[arXiv:1209.3511 [gr-qc]].

\bibitem{Donoghue:2022eay}
J.~F.~Donoghue,
[arXiv:2211.09902 [hep-th]].

\bibitem{Tan:2021lza}
L.~Tan, N.~C.~Tsamis and R.~P.~Woodard,
Phil. Trans. Roy. Soc. Lond. A \textbf{380}, 0187 (2021)
doi:10.1098/rsta.2021.0187
[arXiv:2107.13905 [gr-qc]].

\bibitem{Tan:2022xpn}
L.~Tan, N.~C.~Tsamis and R.~P.~Woodard,
Universe \textbf{8}, no.7, 376 (2022)
doi:10.3390/universe8070376
[arXiv:2206.11467 [gr-qc]].

\bibitem{Miao:2006gj}
S.~P.~Miao and R.~P.~Woodard,
Phys. Rev. D \textbf{74}, 024021 (2006)
doi:10.1103/PhysRevD.74.024021
[arXiv:gr-qc/0603135 [gr-qc]].

\bibitem{Miao:2012bj}
S.~P.~Miao,
Phys. Rev. D \textbf{86}, 104051 (2012)
doi:10.1103/PhysRevD.86.104051
[arXiv:1207.5241 [gr-qc]].

\bibitem{Glavan:2021adm}
D.~Glavan, S.~P.~Miao, T.~Prokopec and R.~P.~Woodard,
JHEP \textbf{03}, 088 (2022)
doi:10.1007/JHEP03(2022)088
[arXiv:2112.00959 [gr-qc]].

\bibitem{Glavan:2013jca}
D.~Glavan, S.~P.~Miao, T.~Prokopec and R.~P.~Woodard,
Class. Quant. Grav. \textbf{31}, 175002 (2014)
doi:10.1088/0264-9381/31/17/175002
[arXiv:1308.3453 [gr-qc]].

\bibitem{Wang:2014tza}
C.~L.~Wang and R.~P.~Woodard,
Phys. Rev. D \textbf{91}, no.12, 124054 (2015)
doi:10.1103/PhysRevD.91.124054
[arXiv:1408.1448 [gr-qc]].

\bibitem{Tsamis:2005hd}
N.~C.~Tsamis and R.~P.~Woodard,
Nucl. Phys. B \textbf{724}, 295-328 (2005)
doi:10.1016/j.nuclphysb.2005.06.031
[arXiv:gr-qc/0505115 [gr-qc]].

\bibitem{Kitamoto:2010et}
H.~Kitamoto and Y.~Kitazawa,
Phys. Rev. D \textbf{83}, 104043 (2011)
doi:10.1103/PhysRevD.83.104043
[arXiv:1012.5930 [hep-th]].

\bibitem{Kitamoto:2011yx}
H.~Kitamoto and Y.~Kitazawa,
Phys. Rev. D \textbf{85}, 044062 (2012)
doi:10.1103/PhysRevD.85.044062
[arXiv:1109.4892 [hep-th]].

\bibitem{Kitamoto:2018dek}
H.~Kitamoto,
Phys. Rev. D \textbf{100}, no.2, 025020 (2019)
doi:10.1103/PhysRevD.100.025020
[arXiv:1811.01830 [hep-th]].

\bibitem{Miao:2021gic}
S.~P.~Miao, N.~C.~Tsamis and R.~P.~Woodard,
JHEP \textbf{03}, 069 (2022)
doi:10.1007/JHEP03(2022)069
[arXiv:2110.08715 [gr-qc]].

\bibitem{Woodard:2023rqo}
R.~P.~Woodard and B.~Yesilyurt,
[arXiv:2302.11528 [gr-qc]].

\bibitem{Starobinsky:1986fx}
A.~A.~Starobinsky,
Lect. Notes Phys. \textbf{246}, 107-126 (1986)
doi:10.1007/3-540-16452-9\_6

\bibitem{Starobinsky:1994bd}
A.~A.~Starobinsky and J.~Yokoyama,
Phys. Rev. D \textbf{50}, 6357-6368 (1994)
doi:10.1103/PhysRevD.50.6357
[arXiv:astro-ph/9407016 [astro-ph]].

\bibitem{Onemli:2002hr}
V.~K.~Onemli and R.~P.~Woodard,
Class. Quant. Grav. \textbf{19}, 4607 (2002)
doi:10.1088/0264-9381/19/17/311
[arXiv:gr-qc/0204065 [gr-qc]].

\bibitem{Onemli:2004mb}
V.~K.~Onemli and R.~P.~Woodard,
Phys. Rev. D \textbf{70}, 107301 (2004)
doi:10.1103/PhysRevD.70.107301
[arXiv:gr-qc/0406098 [gr-qc]].

\bibitem{Kasdagli:2023nzj}
E.~Kasdagli, M.~Ulloa and R.~P.~Woodard,
[arXiv:2302.04808 [gr-qc]].

\bibitem{Glavan:2013mra}
D.~Glavan, T.~Prokopec and V.~Prymidis,
Phys. Rev. D \textbf{89}, no.2, 024024 (2014)
doi:10.1103/PhysRevD.89.024024
[arXiv:1308.5954 [gr-qc]].

\bibitem{Glavan:2014uga}
D.~Glavan, T.~Prokopec and D.~C.~van der Woude,
Phys. Rev. D \textbf{91}, no.2, 024014 (2015)
doi:10.1103/PhysRevD.91.024014
[arXiv:1408.4705 [gr-qc]].

\bibitem{Glavan:2015cut}
D.~Glavan, T.~Prokopec and T.~Takahashi,
Phys. Rev. D \textbf{94}, 084053 (2016)
doi:10.1103/PhysRevD.94.084053
[arXiv:1512.05329 [gr-qc]].

\bibitem{Glavan:2017jye}
D.~Glavan, T.~Prokopec and A.~A.~Starobinsky,
Eur. Phys. J. C \textbf{78}, no.5, 371 (2018)
doi:10.1140/epjc/s10052-018-5862-5
[arXiv:1710.07824 [astro-ph.CO]].

\bibitem{Belgacem:2021ieb}
E.~Belgacem and T.~Prokopec,
Phys. Lett. B \textbf{831}, 137174 (2022)
doi:10.1016/j.physletb.2022.137174
[arXiv:2111.04803 [astro-ph.CO]].

\bibitem{Vedder:2022spt}
C.~J.~G.~Vedder, E.~Belgacem, N.~E.~Chisari and T.~Prokopec,
JCAP \textbf{03}, 016 (2023)
doi:10.1088/1475-7516/2023/03/016
[arXiv:2209.00440 [astro-ph.CO]].

\bibitem{Dolgov:2005se}
A.~Dolgov and D.~N.~Pelliccia,
Nucl. Phys. B \textbf{734}, 208-219 (2006)
doi:10.1016/j.nuclphysb.2005.12.002
[arXiv:hep-th/0502197 [hep-th]].

\bibitem{Itzykson:1980rh}
C.~Itzykson and J.~B.~Zuber,
McGraw-Hill, 1980,
ISBN 978-0-486-44568-7

\bibitem{Weinberg:1996kr}
S.~Weinberg,
Cambridge University Press, 2013,
ISBN 978-1-139-63247-8, 978-0-521-67054-8, 978-0-521-55002-4
doi:10.1017/CBO9781139644174

\bibitem{Planck:2018jri}
Y.~Akrami \textit{et al.} [Planck],
Astron. Astrophys. \textbf{641}, A10 (2020)
doi:10.1051/0004-6361/201833887
[arXiv:1807.06211 [astro-ph.CO]].

\bibitem{BICEP:2021xfz}
P.~A.~R.~Ade \textit{et al.} [BICEP and Keck],
Phys. Rev. Lett. \textbf{127}, no.15, 151301 (2021)
doi:10.1103/PhysRevLett.127.151301
[arXiv:2110.00483 [astro-ph.CO]].

\bibitem{Starobinsky:1980te}
A.~A.~Starobinsky,
Phys. Lett. B \textbf{91}, 99-102 (1980)
doi:10.1016/0370-2693(80)90670-X

\bibitem{Planck:2018vyg}
N.~Aghanim \textit{et al.} [Planck],
Astron. Astrophys. \textbf{641}, A6 (2020)
[erratum: Astron. Astrophys. \textbf{652}, C4 (2021)]
doi:10.1051/0004-6361/201833910
[arXiv:1807.06209 [astro-ph.CO]].

\end{thebibliography}
\end{document}